\documentstyle[eqsecnum,aps]{revtex}

\draft

\begin{document}

\title{Incomplete nonextensive statistics and zeroth law of thermodynamics}

\author{Qiuping A. Wang}
\address{Institut Sup\'erieur des Mat\'eriaux du Mans,\\ 44, Avenue F.A.
Bartholdi, 72000 Le Mans, France}

\date{\today}

\maketitle

\begin{abstract}
We show that the zeroth law of thermodynamics holds within an
alternative version of nonextensive statistical mechanics based
on {\it incomplete probability distribution}. This generalized
zeroth law leads to a generalized definition of thermodynamic
functions which are possible to be used for systems with
important nonextensivity (nonadditivity) in energy, volume or
other external variables.
\end{abstract}


\newpage

\section{Introduction}
In its most recent version\cite{Tsal98}, Tsallis nonextensive
statistical mechanics is constructed on the basis of following
three postulates : the nonextensive entropy

\begin{equation}                                    \label{1}
S_q= -k\frac{1-\sum_ip_i^q}{1-q} , (q \in R)
\end{equation}
where $p_i$ is the probability that the system of interest is at
the state $i$, the probability normalization
\begin{equation}                                    \label{2}
\sum_ip_i=1.
\end{equation}
and the expectation of internal energy $U_q$ of the system :
\begin{equation}                                    \label{3}
U_q=\frac{\sum_i p_i^qE_i}{\sum_ip_i^q}.
\end{equation}
where $E_i$ is the value of hamiltonian $\hat{H}$ at the state
$i$ of the system. It is worth emphasizing that this formalism,
as other current probabilistic theories, is based on the
normalization for {\it complete probability distributions}, i.e.
the sum over $i$ in Eq.(\ref{2}) is to be carried out over all the
possible states of the system of interest.

The above three equations lead to following nonextensivity in
entropy for a isolated system composed of two subsystems $A$ and
$B$ verifying the factorization of joint probability
$p_{ij}(A+B)=p_i(A)p_j(B)$ :
\begin{equation}                                    \label{4}
S_q(A+B)=S_q(A)+S_q(B)-\frac{q-1}{k}S_q(A)S_q(B)
\end{equation}

Recently, the zeroth law of thermodynamics in this formalism
becomes a centre of scientific interest. It was believed that this
law could not exist in the framework of nonextensive statistical
mechanics\cite{Raggio}. If this was true, the definition of heat
and work would be impossible and the application of this
statistical mechanics to thermodynamic systems would be severely
disturbed. On the basis of Eq.(\ref{4}) and the assumption of
extensive energy $U_q(A+B)=U_q(A)+U_q(B)$, Abe and
co-workers\cite{Abe99,Abe01,Mart00,Mart01,Tora01} discussed the
establishment of the zeroth law of thermodynamics within Tsallis
theory and obtain
\begin{equation}                                    \label{10}
\frac{\partial S_q(A)}{\partial U_q(A)}Z_q^{q-1}(A)=\frac{\partial
S_q(B)}{\partial U_q(B)}Z_q^{q-1}(B)
\end{equation}
or
\begin{equation}                                    \label{11}
T(A)=T(B)
\end{equation}
where $1/T=\frac{\partial S_q}{\partial U_q}Z_q^{q-1}$ is defined
as the measurable absolute temperature. One of the consequences
of this approach is that the Clausius definition of entropy is
modified and written as

\begin{equation}                                    \label{11a}
dS_q=Z^{1-q}\frac{dQ}{T}
\end{equation}
where $dQ$ is of course the heat transfer from the surroundings
to the system of interest. This modification has in turn other
consequences on the definition of Helmoholtz free energy and
generalized forces. The reader is referred to the reference
\cite{Abe01} for detailed discussion on this topic.

In this paper, I show an attempt to establish the zeroth law
within an alternative formalism of nonextensive
statistics\cite{Wang00} without neglecting the nonextensivity in
energy. This formalism is based on an information theory related
to {\it incomplete probability distribution} \cite{Reny66} with
which Eq. (\ref{2}) does not hold because
\begin{equation}                                            \label{2b}
\sum_{i=1}^{v}p_i=Q\neq 1
\end{equation}
where $v$ is only the number of the accessible or known states of
the system of interest and so can be greater or smaller than the
total number of all states of the system. This situation can
occur if, for example, we do not know all the interactions in the
system and have to neglect the unknown interaction in the
hamiltonian and in the equation of motion. The solution of this
equation should give us the accessible states.

\section{Incomplete normalization}
When it is impossible to count all the possible states or to know
the exact probability distribution of a system, the probabilities
become nonadditive and do not sum to one. The necessity of the
introduction of this kind of nonadditive probabilities was first
noticed by economists\footnote{I would like to cite some comments
of economists on this subject. J. Dow, M.H. Simonsen and S.
Werlang wrote\cite{Dow}, concerning the financial risk and
uncertainty : "With a nonadditive probability measure, the
'probability' that either of two mutually exclusive events will
occur is not necessarily equal to the sum of their two
'probabilities'. If it is less than the sum, the expected-utility
calculations using this probability measure will reflect
uncertainty aversion as well as (possibly) risk aversion. The
reader may be disturbed by 'probabilities' that do not sum to
one. It should be stressed that the probabilities, together with
the utility function, provide a representation of behavior. They
are not 'objective probabilities'". "Uncertainty means, in fact,
incomplete information about the true probabilities ... The
attractiveness of the concept of sub-additive probabilities is
that it might provide the best possible description for what is
behind the widespread notion of 'subjective probabilities' in the
theory of financial decisions." It is also of interest to note
that, in the last decade, there was a strong development of the
method of Backward Stochastic Differential Equation with a
nonextensive q-expectation which may be applied to describe the
"risk aversion" behavior of financial decisions \cite{Peng}}.
Recently, in order to address this kind of probabilities, we
proposed to replace Eqs. (\ref{2}) and (\ref{3}) with following
relations \cite{Wang00} :

\begin{equation}                                            \label{2a}
\sum_i^v\frac{p_i}{Q}=\sum_{i=1}^{v}p_i^q=1, (q\in[0,\infty])
\end{equation}
and
\begin{equation}                                            \label{3a}
U_q=\sum_{i=1}^{v}p_i^qE_i.
\end{equation}
We would like to indicate here that the parameter $q$ is in this
way logically related to the quantity $Q$ in Eq.(\ref{2b}) and so
to the interactions neglected in the hamiltonian of the system.
For example, for a micro-canonical ensemble, we can write :
$q=1-\frac{lnQ}{lnp}$ or $Q=v^{\frac{q-1}{q}}$. This relation may
help us to understand $q$ values different from unity obtained for
complex systems.

On the basis of Eqs.(\ref{2a}) and (\ref{3a}), considering the
usual postulates of Shannon information theory and an information
measure $I$ with generalized Hartley formula
($I(x)=\frac{x^{1-q}-1}{1-q}$), we obtain following entropy for
nonextensive systems :
\begin{equation}                                        \label{1b}
S_q=-k\sum_{i=1}^{v}p_i^q\frac{p_i^{1-q}-1}{1-q}
=-k\frac{\sum_{i=1}^{v}p_i-\sum_{i=1}^{v}p_i^q}{1-q}
\end{equation}
which is just Tsallis entropy Eq(\ref{1}) provided that the
normalization Eq.(\ref{2}) holds. If we consider the incomplete
normalization Eq.(\ref{2a}) and expectation Eq.(\ref{3a}), this
entropy leads to different nonextensive properties which is
consistent with following assumptions concerning an isolated
system composed of two correlated systems $A$ and
$B$\cite{Wang00} :

\begin{equation}                                    \label{4a}
S_q(A+B)=S_q(A)+S_q(B)+\frac{q-1}{k}S_q(A)S_q(B).
\end{equation}
and for internal energy :
\begin{equation}                                    \label{9a}
U_q(A+B)=U_q(A)+U_q(B)+(q-1)\beta U_q(A)U_q(B).
\end{equation}
These relations should be considered as two fundamental
hypotheses of the incomplete statistics on the basis of which the
zeroth law holds.

\section{Zeroth law of thermodynamics}

From Eq. (\ref{4a}), a small variation of the total entropy can
be written as :

\begin{eqnarray}                                    \label{37}
\delta S_q(A+B) & = & [1+\frac{q-1}{k}S_q(B)]\delta S_q(A)+
[1+\frac{q-1}{k}S_q(A)]\delta S_q(B) \\ \nonumber
    & = & [1+\frac{q-1}{k}S_q(B)]
\frac{\partial S_q(A)}{\partial U_q(A)} \delta U_q(A) +
[1+\frac{q-1}{k}S_q(A)] \frac{\partial S_q(B)}{\partial U_q(B)}
\delta U_q(B).
\end{eqnarray}
From Eq. (\ref{9a}), the variation of the total internal energy
is given by :

\begin{eqnarray}                                    \label{38}
\delta U_q(A+B) & = & [1+(q-1)\beta U_q(B)]\delta U_q(A)+
[1+(q-1)\beta U_q(A)]\delta U_q(B).
\end{eqnarray}
It is supposed that the total system $(A+B)$ is completely
isolated. So $\delta U_q(A+B)=0$ which leads to :
\begin{eqnarray}                                    \label{39}
\frac{\delta U_q(A)}{1+(q-1)\beta U_q(A)}= -\frac{\delta
U_q(B)}{1+(q-1)\beta U_q(B)}
\end{eqnarray}
At the equilibrium of the composite system $(A+B)$, we have
$\delta S_q(A+B)=0$. So Eqs. (\ref{37}) and (\ref{39}) lead to :

\begin{eqnarray}                                    \label{40}
\frac{1+(q-1)\beta U_q(A)}{1+\frac{q-1}{k}S_q(A)}\frac{\partial
S_q(A)}{\partial U_q(A)}= \frac{1+(q-1)\beta
U_q(B)}{1+\frac{q-1}{k}S_q(B)}\frac{\partial S_q(B)}{\partial
U_q(B)}.
\end{eqnarray}
Now with the help of the incomplete distribution function given
in reference
\cite{Wang00}\footnote{$p_i=\frac{1}{Z_q}[1-(1-q)\beta
E_i]^{\frac{1}{1-q}}$ with $Z_q^q=\sum_i^v[1-(1-q)\beta
E_i]^{\frac{q}{1-q}}$.}, we easily obtain
\begin{eqnarray}                                    \label{40a}
\sum_ip_i=\frac{1}{Z_q^{1-q}}[1+(q-1)\beta U_q].
\end{eqnarray}
From Tsallis entropy Eq.(\ref{1b}) and the incomplete
normalization Eq.(\ref{2a}), we get
\begin{eqnarray}                                    \label{40b}
\sum_ip_i=1+\frac{q-1}{k}S_q.
\end{eqnarray}
Eqs.(\ref{40a}) and (\ref{40b}) lead to
\begin{eqnarray}                                    \label{41}
\frac{1+(q-1)\beta U_q}{1+\frac{q-1}{k}S_q}=Z_q^{1-q}
\end{eqnarray}
which recasts Eq.(\ref{40}) as follows :
\begin{eqnarray}                                    \label{42}
Z_q^{1-q}(A)\frac{\partial S_q(A)}{\partial U_q(A)}=
Z_q^{1-q}(B)\frac{\partial S_q(B)}{\partial U_q(B)}.
\end{eqnarray}
Eq.(\ref{41}) also means
\begin{eqnarray}                                    \label{43}
Z_q^{1-q}\frac{\partial S_q}{\partial U_q}=\beta.
\end{eqnarray}
or
\begin{eqnarray}                                    \label{43a}
\frac{\partial \textbf{S}_q}{\partial U_q}=\beta.
\end{eqnarray}
where $\textbf{S}_q=Z_q^{1-q}S_q$. We finally have
\begin{eqnarray}                                    \label{44}
\beta (A)=\beta (B)
\end{eqnarray}
where $\beta$ is to be considered as the generalized inverse
temperature $1/kT$. Eq.(\ref{44}) is the generalized zeroth law of
thermodynamics for nonextensive systems with the same parameter
$q$. It suggests that the Clausius' definition of thermodynamic
entropy has to be modified. Now we should write
\begin{eqnarray}                                    \label{45}
dQ=Td\textbf{S}_q.
\end{eqnarray}
So the first law of thermodynamics should be written as follows
\begin{eqnarray}                                 \label{e14}
dU_q=Td\textbf{S}_q+YdX.
\end{eqnarray}
where $Y$ is the generalized force and $X$ the correspondent
external variable. The free-energy $F_q$ should be defined as
\begin{eqnarray}                                    \label{e15}
dF_q=-\textbf{S}_qdT+YdX
\end{eqnarray}
or
\begin{eqnarray}                                    \label{e16}
F_q=U_q-T\textbf{S}_q,
\end{eqnarray}
which leads to, with the help of Eq.(\ref{41}) :
\begin{equation}
F_q=-kT\frac{Z^{1-q}-1}{1-q}.                         \label{e17}
\end{equation}
We can easily show that $Z_q(C)=Z_q(A)Z_q(B)$. So for a composite
system $C=A+B$, we can write :
\begin{eqnarray}                                    \label{e17a}
F_q(C)&=& F_q(A)+F_q(B)-(1-q)\beta F_q(A)F_q(B).
\end{eqnarray}
We also have, for the heat capacity
\begin{equation}                                        \label{e18}
C_X=\{ \frac{dQ}{dT} \} _X =T \{ \frac{\partial
\textbf{S}_q}{\partial T} \} _X=-T \{ \frac{\partial^2
F_q}{\partial T^2} \}_X
\end{equation}
and for the equation of state :
\begin{equation}                                      \label{e18a}
Y=\{ \frac{\partial F_q}{\partial X} \} _T
\end{equation}

We can also address other ensembles like, for example,
isothermal-isobaric ensemble. The Gibbs free energy $G$ in this
case is given by
\begin{eqnarray}                                    \label{e19}
G_q=U_q-T\textbf{S}_q+pV=-kT\frac{Z^{1-q}-1}{1-q},
\end{eqnarray}
where $Z_q^q=\sum_i^v[1-(1-q)\beta (E_i+pV_i)]^{\frac{q}{1-q}}$,
$p$ is the pressure, $V_i$ the volume of the system at state $i$
(with $V$ as average value). We naturally have
\begin{eqnarray}                                    \label{e20}
G_q(C)&=& G_q(A)+G_q(B)-(1-q)\beta G_q(A)G_q(B).
\end{eqnarray}
and
\begin{equation}                                     \label{e21}
V=\{ \frac{\partial G_q}{\partial p} \} _T.
\end{equation}

\section{Conclusion}
It is shown that the zeroth law of thermodynamics can hold within
the nonextensive statistics based on the incomplete normalization
without neglecting the nonextensivity in energy or other physical
quantities. So this nonextensive statistical mechanics offers the
possibility to address systems in which the nonadditivity in
energy, volume or other external variables can not be neglected.

\end{document}